%% file: paper.tex
\documentclass[sigconf,10pt,preprint]{acmart}

\input{includes/packages}
\input{includes/settings}

\if\showcomments1
    \usepackage[textsize=tiny]{todonotes}
\else
    \usepackage[disable]{todonotes}
\fi

\begin{document}

%%
%% The "title" command has an optional parameter,
%% allowing the author to define a "short title" to be used in page headers.
\title{\name{}: Interactive Serverless Compute on Programmable SmartNICs}

%%
%% The "author" command and its associated commands are used to define
%% the authors and their affiliations.
%% Of note is the shared affiliation of the first two authors, and the
%% "authornote" and "authornotemark" commands
%% used to denote shared contribution to the research.
%\author{Ben Trovato}
%\authornote{Both authors contributed equally to this research.}
%\email{trovato@corporation.com}
%\orcid{1234-5678-9012}
%\author{G.K.M. Tobin}
%\authornotemark[1]
%\email{webmaster@marysville-ohio.com}
%\affiliation{%
%  \institution{Institute for Clarity in Documentation}
%  \streetaddress{P.O. Box 1212}
%  \city{Dublin}
%  \state{Ohio}
%  \postcode{43017-6221}
%}

\author{Sean Choi, Muhammad Shahbaz, Balaji Prabhakar, and Mendel Rosenblum}
\affiliation{%
  \institution{{\em Stanford University}}
%  \streetaddress{1 Th{\o}rv{\"a}ld Circle}
%  \city{Hekla}
%  \country{Iceland}
}
%\email{larst@affiliation.org}

%\author{Valerie B\'eranger}
%\affiliation{%
%  \institution{Inria Paris-Rocquencourt}
%  \city{Rocquencourt}
%  \country{France}
%}
%
%\author{Aparna Patel}
%\affiliation{%
% \institution{Rajiv Gandhi University}
% \streetaddress{Rono-Hills}
% \city{Doimukh}
% \state{Arunachal Pradesh}
% \country{India}}
%
%\author{Huifen Chan}
%\affiliation{%
%  \institution{Tsinghua University}
%  \streetaddress{30 Shuangqing Rd}
%  \city{Haidian Qu}
%  \state{Beijing Shi}
%  \country{China}}

%\author{Charles Palmer}
%\affiliation{%
%  \institution{Palmer Research Laboratories}
%  \streetaddress{8600 Datapoint Drive}
%  \city{San Antonio}
%  \state{Texas}
%  \postcode{78229}}
%\email{cpalmer@prl.com}
%
%\author{John Smith}
%\affiliation{\institution{The Th{\o}rv{\"a}ld Group}}
%\email{jsmith@affiliation.org}
%
%\author{Julius P. Kumquat}
%\affiliation{\institution{The Kumquat Consortium}}
%\email{jpkumquat@consortium.net}

%\author{Paper \#42}
%\affiliation{\institution{\pageref{lastpage} Pages Body, \pageref{totalpage} Pages Total}}
%\email{jsmith@affiliation.org}

%%
%% By default, the full list of authors will be used in the page
%% headers. Often, this list is too long, and will overlap
%% other information printed in the page headers. This command allows
%% the author to define a more concise list
%% of authors' names for this purpose.
%\renewcommand{\shortauthors}{Trovato and Tobin, et al.}

%%
%% Setup todo notes for comments and annotations
\if\showcomments1
    \setcounter{page}{0}
    \listoftodos {}
    \clearpage
\fi

%%
%% The abstract is a short summary of the work to be presented in the
%% article.
\input{text/abstract}

\maketitle

%% Removes running page headers
\if\removepageheaders1
    \pagestyle{plain}
\fi

\input{text/introduction}
\input{text/background}

\input{text/motivation}
\input{text/overview}

\input{text/implementation}

\input{text/evaluation}
\input{text/discussion}
\input{text/related-work}

\balance\input{text/conclusion}

\label{lastpage}

%%
%% The acknowledgments section is defined using the "acks" environment
%% (and NOT an unnumbered section). This ensures the proper
%% identification of the section in the article metadata, and the
%% consistent spelling of the heading.
%\begin{acks}
%To Robert, for the bagels and explaining CMYK and color spaces.
%\end{acks}

%%
%% The next two lines define the bibliography style to be used, and
%% the bibliography file.
\bibliographystyle{ACM-Reference-Format}
\bibliography{paper}

\label{totalpage}

%%
%% If your work has an appendix, this is the place to put it.
%\appendix

%\section{Research Methods}

\end{document}

%% file: includes/packages.tex
\usepackage{xspace} 
\usepackage{enumitem} % left margin for itemize
\usepackage{color, colortbl}
\usepackage{xcolor}
\usepackage{textgreek}

\usepackage[capitalise]{cleveref}

\usepackage{siunitx}

\usepackage{listings}
\usepackage[utf8]{inputenc}

\usepackage{booktabs}  
\usepackage{array}
\usepackage{multirow}
\usepackage{diagbox}

\usepackage{balance}
\usepackage{xurl}

\usepackage[multiple]{footmisc}

%% file: includes/settings.tex
\def\setuppreprint{1}
\def\setupcameready{0}

\if\setupcameready1
    \setcopyright{acmcopyright}
    \copyrightyear{2018}
    \acmYear{2018}
    \acmDOI{10.1145/1122445.1122456}
\else
    \setcopyright{none}
    \acmDOI{}
\fi

\if\setupcameready1
    \acmConference[Woodstock '18]{Woodstock '18: ACM Symposium on Neural
        Gaze Detection}{June 03--05, 2018}{Woodstock, NY}
    \acmBooktitle{Woodstock '18: ACM Symposium on Neural Gaze Detection,
        June 03--05, 2018, Woodstock, NY}
    \acmPrice{15.00}
    \acmISBN{978-1-4503-9999-9/18/06}
\else
    \acmISBN{}
\fi

\if\setupcameready1
    \acmSubmissionID{123-A56-BU3}
\fi

\if\setupcameready1
\else
\fi

\if\setupcameready0
    \renewcommand\footnotetextcopyrightpermission[1]{} 
\fi

\if\setupcameready1
    \def\removepageheaders{0} 
\else
    \def\removepageheaders{1}
\fi

\newcommand{\ie}{{\em i.e.}}
\newcommand{\eg}{{\em e.g.}}

\newcommand{\name}{$\lambda$-NIC\xspace}
\newcommand{\matchlambda}{Match+Lambda}

\if\setupcameready1
    \def\showcomments{0}
\else
    \if\setuppreprint1
        \def\showcomments{0}
    \else
        \def\showcomments{1}
    \fi
\fi

\makeatletter
\def\@parfont{\bfseries\itshape}
\def\@subsubsecfont{\sffamily\bfseries\section@raggedright}
\makeatother

\definecolor{codegreen}{rgb}{0,0.6,0}
\definecolor{codegray}{rgb}{0.5,0.5,0.5}
\definecolor{codepurple}{rgb}{0.58,0,0.82}
\definecolor{backcolour}{rgb}{0.95,0.95,0.92}
 
\lstdefinestyle{match_lambda_style}{
    backgroundcolor=\color{backcolour},   
    commentstyle=\color{codegreen},
    keywordstyle=\color{magenta},
    numberstyle=\tiny\color{codegray},
    stringstyle=\color{codepurple},
    basicstyle=\ttfamily\footnotesize,
    breakatwhitespace=false,         
    breaklines=true,                 
    captionpos=b,                    
    keepspaces=true,                 
    numbers=left,                    
    numbersep=5pt,                  
    showspaces=false,                
    showstringspaces=false,
    showtabs=false,                  
    tabsize=2
}
 

%% file: text/abstract.tex
\begin{abstract}
There is a growing interest in serverless compute, a cloud computing model that automates infrastructure resource-allocation and management while billing customers only for the resources they use. 
Workloads like stream processing benefit from high elasticity and fine-grain pricing of these serverless frameworks. 
However, so far, limited concurrency and high latency of server CPUs prohibit many interactive workloads (\eg, web servers and database clients) from taking advantage of serverless compute to achieve high performance.

In this paper, we argue that server CPUs are ill-suited to run serverless workloads (\ie, lambdas) and present \name{}, an open-source framework, that runs interactive workloads directly on a SmartNIC; more specifically an ASIC-based NIC that consists of a dense grid of Network Processing Unit (NPU) cores. 
\name{} leverages SmartNIC's proximity to the network and a vast array of NPU cores to simultaneously run thousands of lambdas on a single NIC with strict tail-latency guarantees. 
To ease development and deployment of lambdas, \name{} exposes an event-based programming abstraction, {\em \matchlambda{}}, and a machine model that allows developers to compose and execute lambdas on SmartNICs easily. 
Our evaluation shows that \name{} achieves up to 880x and 736x improvements in workloads' response latency and throughput, respectively, while significantly reducing host CPU and memory usage.
\end{abstract}

%% file: text/introduction.tex
\section{Introduction}
\label{sec:introduction}

Serverless compute is emerging as an attractive cloud computing model that lets developers focus only on the core applications---building the workloads as small, fine-grained custom programs (\ie, lambdas)---without having to worry about the infrastructure they run on. 
Cloud providers dynamically provision, deploy, patch, and monitor the infrastructure and its resources (\eg, compute, storage, memory, and network) for these workloads; with tenants only paying for the resources they consume at millisecond increments. 
Serverless compute lowers the barrier to entry, especially, for organizations lacking expertise, manpower, and budget to efficiently manage the underlying infrastructure resources.

Today, all major cloud vendors offer some form of serverless frameworks, such as Amazon Lambda~\cite{amzn_lmda_wp}, Google Cloud Functions~\cite{goog_func}, and Microsoft Azure Functions~\cite{msft_func}, along with open-source versions like OpenFaaS~\cite{openfaas} and OpenWhisk~\cite{openwhisk}. 
These frameworks rely on server virtualization (\ie, virtual machines (VMs)~\cite{vm}) and container technologies (\ie, Docker~\cite{docker}) to execute and scale tenants' workloads. 
These technologies are designed to maximize utilization of the providers' physical infrastructure, while presenting each tenant with its own view of a completely isolated machine. 
However, in serverless compute, where server management is hidden from tenants, these virtualization technologies become redundant, unnecessarily bloating the code-size of serverless workloads, and causing processing delays (of hundreds of milliseconds) and memory overheads (of tens of megabytes)~\cite{7899408}. 
The increased overheads also limit the concurrent execution (less than hundred or so) of these workloads on a single server, hence, raising the overall cost of running such workloads in a data center. 

The cloud-computing industry is now realizing these issues and some providers, such as Google and CloudFlare, have already started developing alternative frameworks (like Isolate~\cite{cloudflare_isolate}) that remove these technology layers (\eg, containers) and run serverless workloads directly on a bare-metal server~\cite{cloudflare_isolate}. 
However, these bare-metal alternatives are inherently limited by the design restrictions of CPU-based architectures, which exacerbate when running at the scale of cloud data centers~\cite{187030}. 
CPUs are designed to process sequence of instructions (\ie, a function) blazingly fast. 
They are not designed to run thousands of such small, discrete functions in parallel---a typical server CPU has in the order of 4 to 28 cores that can run up to 56 simultaneous threads~\cite{intel_platinum}. 
Each serverless function interrupts a CPU core to store the state (\eg, registers and memory) of the currently running function and load itself with the new one, resulting in wasting tens of milliseconds worth of CPU cycles per context switch (such wasted cycles increase the overall costs for the cloud providers~\cite{Firestone:2018:AAN:3307441.3307446}).
Thus, with ever increasing network speeds---100/400G NICs are on the horizon---these overheads quickly add up, limiting throughput and leading to long-tail latency in the order of 100s of milliseconds~\cite{Kaufmann:2016:HPP:2954679.2872367}.

Recently, public cloud providers are deploying SmartNICs in an attempt to reduce load on host CPUs~\cite{Firestone:2018:AAN:3307441.3307446}. 
So far, these attempts have been limited to offloading ad-hoc application tasks (like TCP offload, VXLAN tunneling, overlay networking, or some partial computation) to accelerate network processing of the hosts~\cite{floem, liu2019e3, Liu:2019:ODA:3341302.3342079}.
However, modern SmartNICs, more specifically ASIC-based NICs, consist of hundreds of RISC processors (\ie, NPUs)~\cite{netronome_cx}, each with their local instruction store and memory.
These SmartNICs are more flexible and can run many discrete functions in parallel at high speed and low latency---unlike GPUs and FPGAs, which are optimized to accelerate specific workloads~\cite{DBLP:journals/corr/BahrampourRSS15, Firestone:2018:AAN:3307441.3307446, Putnam:2014:RFA:2665671.2665678},

Serverless workloads by design are small, presenting unique opportunities for SmartNICs to accelerate them, while also achieving strict tail-latency guarantees. 
However, the main shortcomings of using SmartNICs come from their programming complexity.
Programming SmartNICs is a formidable undertaking that requires intimate familiarity with NIC's system and resource architecture (\eg, memory hierarchy, and multi-core parallelism and pipelining). 
Developers need to carefully partition, schedule, and arbitrate these resources to maximize performance of their applications, which is a characteristic that is counter to the motivation behind serverless compute (\ie, where developers are unaware of the architectural details of the underlying infrastructure).
Furthermore, each application has to explicitly handle packet processing as there is no notion of a network stack in SmartNICs.

In this paper, we present \name{}, a framework for running interactive serverless workloads entirely on SmartNICs.
\name{} supports a new programming abstraction (\matchlambda{}) along with a machine model---an extension of P4's match-action abstraction with more sophisticated actions---and helps address the shortcomings of SmartNICs in five key ways.
First, users provide their lambdas, which \name{} compiles and then, at runtime, selects to execute by matching on the header of the incoming requests' packets. 
Second, users program their lambdas assuming a flat memory model; \name{} analyzes the memory-access patterns (\ie, read, write, or both) and sizes, and optimally maps these lambdas across different memory hierarchies of the NICs while ensuring that memory accesses are isolated.
Third, \name{} infers which packet headers are used by each lambda and automatically generates the corresponding parser for the headers, thus eliminating the need for manually specifying packet-processing logic within these lambdas.
%Furthermore, \name{} coalesces the packet-parsing logic to reduce the program size, allowing more lambdas to fit on a single NPU.
Fourth, instead of partitioning and scheduling a single lambda across multiple NPUs, \name{} assumes a run-to-completion (RTC) model exploiting the fact that lambdas are small and can run within a single NPU.
The vast array of NPU cores and short service times of lambdas further mitigate head-of-line-blocking and performance issues that lead to high tail latency.
Lastly, serverless functions mostly communicate using independent, mutually-exclusive request-response pairs and do not need the strict in-order, streaming delivery semantic provided by TCP~\cite{Kogias:2019:RMR:3358807.3358881}. 
\name{}, therefore, employs a weakly-consistent delivery semantic~\cite{Caulfield:2016:CAA:3195638.3195647, Kogias:2019:RMR:3358807.3358881}, alongside RDMA~\cite{rdma}, for communication between serverless workloads---processing requests directly within the NIC cores without involving the host CPU. In summary, we make the following contributions:

\begin{itemize}[leftmargin=*]
    \item We introduce a new abstraction for our \name{} framework, called \matchlambda{} (\S\ref{subsec:match_lambda}), and a machine model (\S\ref{subsec:runtime}) to easily and effectively code and execute lambdas on modern SmartNICs. 
    \item We develop an open-source implementation of \name{} using P4-enabled Netronome SmartNICs~(\S\ref{sec:implementation}), and implement methodologies to optimize lambdas to efficiently utilize the SmartNIC resources (\S\ref{subsec:compiler_optimization}).
    \item We evaluate \name{} and show up to two orders of magnitude improvement in latency and throughput compared to existing serverless compute frameworks (\S\ref{sec:evaluation}).
\end{itemize}

We begin with a background on the current state-of-the-art in cloud-computing frameworks and SmartNICs (\S\ref{sec:background}) followed by the challenges and motivations behind \name{} (\S\ref{sec:motivation}).
We then describe the overall architecture of \name{} (\S\ref{sec:overview}) with an extensive evaluation of the system (\S\ref{sec:evaluation}). 
Finally, we conclude by discussing \name{}'s limitations and future works (\S\ref{sec:discussion}), and comparisons with related implementations (\S\ref{sec:related}).

%% file: text/background.tex
\section{Background}
\label{sec:background}
We now discuss the latest advancements in cloud-computing frameworks and programmable SmartNICs, which are the core building blocks of \name{}.

\subsection{Cloud Computing Frameworks}
\label{subsec:b_cc_env}

\begin{figure}[t]
  \centering
  \includegraphics[width=\linewidth]{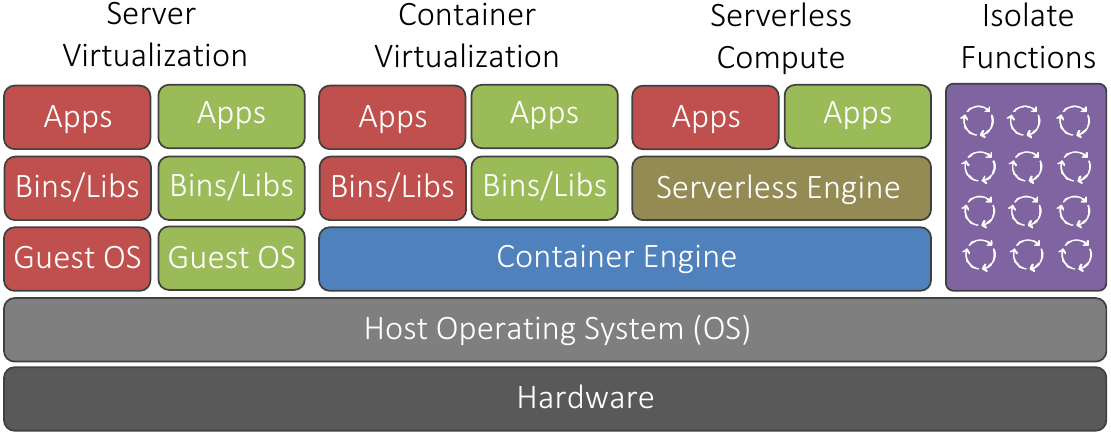}
  \caption{Overview of cloud-computing frameworks and how they partition compute across various layers.}
  \label{fig:cont_vs_vm}
\end{figure}

\Cref{fig:cont_vs_vm} illustrates four different cloud-computing frameworks in use today. 
{\em Server virtualization} is the foremost technology underneath cloud computing that allows a bare-metal server to host multiple VMs each with their independent, isolated environments containing separate Operating Systems (OS), libraries, and applications.
It arrived at a time when advances in hardware made it difficult for a single application to efficiently consume the entire bare-metal server resources, and having multiple applications co-existing on a single server raised various issues (such as isolation and contention for resources).
However, as trends shifted from monoliths to building applications as microservices~\cite{microservice}---to increase manageability, resiliency, and scalability---the overheads of having a separate OS for each microservice were no longer negligible.
This gave rise to {\em container virtualization}~\cite{docker}, which is a way to isolate applications' binaries and libraries while sharing an OS.

\begin{figure}[t]
  \centering
  \includegraphics[width=0.85\linewidth]{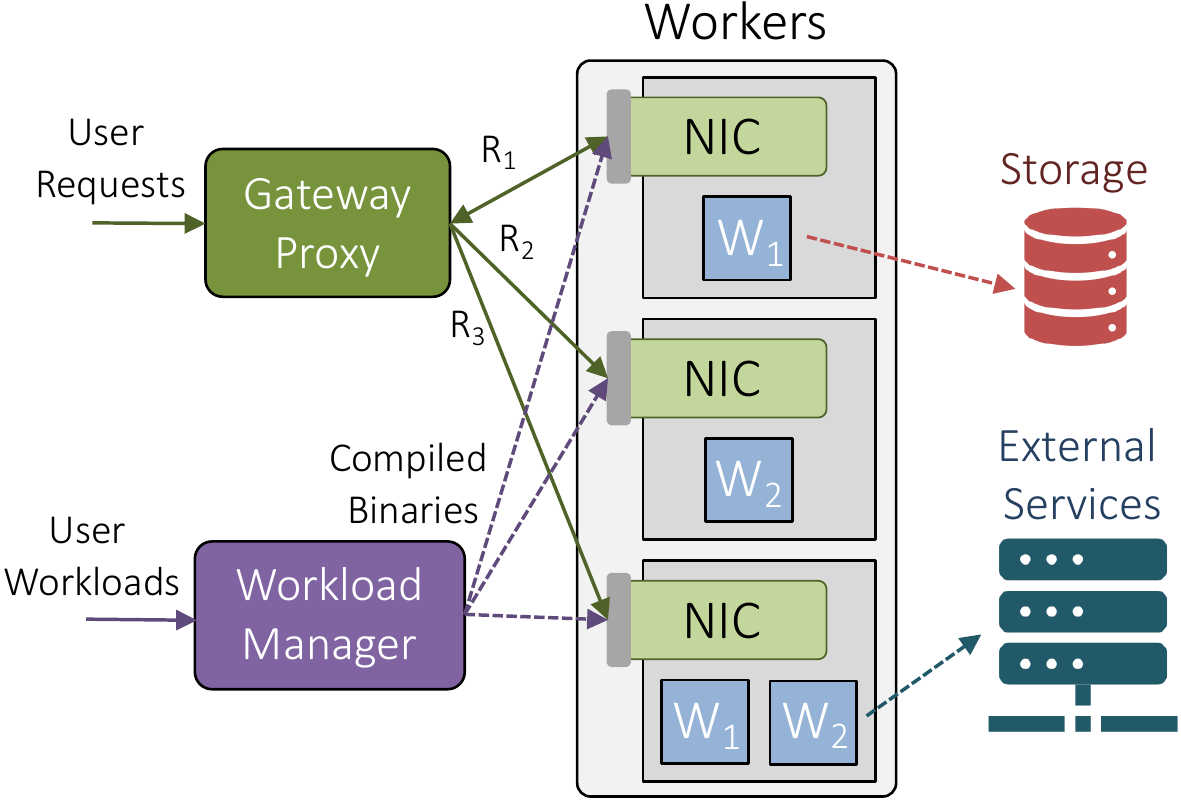}
  \caption{An overview of a general serverless compute framework that executes users' workloads and serves requests to such workloads. ($R_i$ represents the $i$th request to workload $W_i$).}
  \label{fig:serverless_overview}
\end{figure}

Still, with growing complexity and scale of cloud workloads, it became daunting for many users to provision and manage infrastructure resources with tasks requiring fine-grain allocation of resources under changing workload demands.
Early solutions settled on over-provisioning these resources, incurring added cost for idle resources. 
More recently, {\em serverless compute}~\cite{serverless} has emerged as a favorable compute model that alleviates such operational and financial burdens from the users by letting them specify only the workloads, their memory and timing constraints, and when and which events (\eg, API calls, database updates, incoming data streams) to trigger them.
In response, cloud providers independently provision infrastructure resources, deploying a set of containers on-demand to serve workloads' requests.
These containers are quickly taken down once the workloads complete and users are charged only for the time a container is executed.
Serverless workloads are therefore short-lived with strict compute-time and memory limits (up to 15 minutes and \SI{3}{GB}, respectively, for Amazon Lambda~\cite{lambda_limit}). 

\paragraph{Serverless frameworks}
\Cref{fig:serverless_overview} depicts a typical serverless compute framework. 
Workload manager compiles users' workloads to executable binaries, which along with their data dependencies are stored in a global storage (\eg, Amazon S3~\cite{s3}, Google Cloud~\cite{google_cloud_storage}, or Microsoft Azure Storage~\cite{Calder:2011:WAS:2043556.2043571}).
Gateway proxies users' requests (or events) to appropriate workloads, which typically run as containers on a set of dynamically provisioned servers, called worker nodes.
Upon completion, results are written back to the storage or forwarded to other workloads for further execution.

Serverless compute frameworks embed users' workloads within containers managed by orchestration engines (\eg, Docker) running atop an OS, which provide memory, compute, and file-system isolation using OS-based mechanisms (\eg, cgroups~\cite{cgroups} and namespaces~\cite{namespaces}). 
Each container maintains its own libraries and binaries, and communicates with other containers using an overlay network that is set up using virtual switches (like Open vSwitch (OVS)~\cite{ovs}). 

However, running workloads as containers incur additional processing and networking overheads. 
As an alternative, {\em Isolate functions}~\cite{cloudflare_isolate} run workloads directly on the bare-metal server itself (\Cref{fig:cont_vs_vm}), while providing all the benefits of a typical serverless framework (\eg, resource isolation between workloads).
Although still in early development, these Isolate functions are showing promising results: up to 3x improvement in request latency while consuming 90\% less memory than containers with faster startup times.

\paragraph{Serverless workloads}
Today, serverless frameworks find applications in two types of use cases~\cite{azure_serverless_cookbook, lambda_use_cases, goog_cf_usecases}:
(1) Running API backends that serve interactive applications, such as returning static or dynamic content, or key-value responses based on users' requests. 
(2) Processing changes in data stores at real-time, such as cropping a newly uploaded image or running a custom operation on a newly added database entry.

The complexity of lambda functions ranges from running simple arithmetic operations to making machine-learning decisions that are generally short-lived.
A bare-metal server can house thousands of such lambda functions; however, doing so causes CPUs to constantly context switch between these functions.
These, along with other communication overheads, make it difficult for services (using lambda functions) to meet their tail-latency service-level objectives (SLOs)~\cite{cloudflare_isolate}.

To mitigate these issues, efforts are underway to move computation down to SmartNICs; however, unlike \name{}, the focus is mostly on offloading either network processing or some small portion of applications to these NICs~\cite{floem, liu2019e3, Liu:2019:ODA:3341302.3342079}.

\iffalse
\subsection{Programmable Networks}
\label{subsec:b_p_net}
\name{} combines programmable networks and serverless compute. 
Unlike traditional networks, which consist of network devices that run hard-wired programs containing superset of all the desired features, programmable networks consist of switching devices capable of running custom programs~\cite{bosshart13}.
Programmable networks are realized by two recent advances: a programmable network hardware and a language for programming it.
Programmable network hardware has an architecture called RMT~\cite{bosshart13}, consisting of customizable parser, match, and action logic.
To program such hardware, P4~\cite{Bosshart:2014:PPP:2656877.26568900} emerged as a representative high-level language. 
P4 allows easily specifying logic for header parsing, matching, and custom actions written using primitive actions defined in P4 specifications~\cite{p4_spec}.
Some alternate languages include Broadcom's MPL and Berkeley's eBPF instruction set~\cite{eBPF}.
The particular choice of language is typically dictated by hardware vendors and which features the programmers' need.
\fi

\begin{table*}[htp]
\centering
    \begin{tabular}{ p{3cm} | p{4.4cm} | >{\columncolor[gray]{0.9}} p{4.4cm} | p{4.4cm} }
\toprule
                            &
    {\bfseries FPGA-based SmartNICs}  &
    {\bfseries ASIC-based SmartNICs}  &
    {\bfseries SoC-based SmartNICs}   \\
\midrule
    {\bfseries Programmability}                 &
    Hard                                        &
    Limited                       				&
    Easy                                       \\
    {\bfseries Performance}                     &
    10+  cores, low latency                     &
    200+ cores, low latency                     &
    50+ cores, high latency                    \\
    {\bfseries Development cost}&
    High                                        &
    Medium                                      &
    Low                                         \\
%\hline
%    {\bfseries User Cost}                       &
%    Low                                         &
%    Medium                                      &
%    High                                        \\
\bottomrule
\end{tabular}
\captionof{table}{A comparison of various types of SmartNICs.}
\vspace{-25pt}
\label{tab:smartnic-comparisons}
\end{table*}

\subsection{Programmable SmartNICs}
\label{subsec:b_smartnic}

In addition to handling basic networking tasks, SmartNICs can offload more general tasks that a CPU normally handles (\eg, checksum, TCP offload, and more). 
Based on their architecture and processing capabilities, these SmartNICs come in three different types: FPGA-, ASIC-, and SoC-based~\cite{Firestone:2018:AAN:3307441.3307446}. Table~\ref{tab:smartnic-comparisons} summarizes the main differences.

Alongside other issues (\eg, steep development cost and power consumption~\cite{Firestone:2018:AAN:3307441.3307446}), FPGA-based NICs~\cite{fpga} are dominated by the on-chip interconnect overhead~\cite{chen2010optimization} that significantly limits the lookup tables (LUTs) and memory (\eg, SRAM) resources available for executing lambda functions---today's large FPGAs can barely support a small number of processing cores ($< 10$ or so). 
SoC-based NICs (\eg, Mellanox Bluefield~\cite{mellanox_bluefield} and Broadcom Stingray~\cite{broadcom_smartnic}) are easier to program as they run a Linux-like OS on embedded cores (like ARM); however, similar to server CPUs, they are susceptible to high tail latency due to context switch and network stack overheads.
Therefore, it's questionable that these SoC-based NICs can support speeds higher than \SI{100}{Gb}+ with low latency~\cite{Firestone:2018:AAN:3307441.3307446}. 

Due to these limitations of FPGA- and SoC-based NICs, we opted for the ASIC-based NICs when designing \name{}. 
These NICs consist of an ASIC that can sustain traffic rates of \SI{100}{Gbps}+; contain hundreds of non-cache-coherent multi-threaded RISC cores (\eg, NPU, ARM, or RISC-V), operating at \si{GHz} speeds, along with specialized hardware functions (\eg, lookup, load balancing, queuing, and more); and are capable of running embarrassingly parallel workloads with low latency.
Furthermore, recent advances in the design of these SmartNICs (\eg, Netronome Agilio~\cite{netronome_cx} and Marvell LiquidIO~\cite{cavium_liquidio}) make it easier for users to customize the NIC's data-plane logic (\eg, parse, match, and action) using, for example, P4~\cite{Bosshart:2014:PPP:2656877.26568900} and Micro-C~\cite{netronome_programming} programs, thus exposing dataflow and C-like abstractions a typical programmer is familiar with, without the need for an OS.

In~\S\ref{sec:evaluation}, we demonstrate how the unique characteristics of serverless workloads (\ie, short-lived with strict compute and memory limits) make ASIC-based SmartNICs a viable execution platform to accelerate lambda functions.

%% file: text/motivation.tex
\section{Motivation \& Challenges}
\label{sec:motivation}

The key motivation behind \name{} is to accelerate interactive serverless workloads by: (1) mitigating excessive compute and network virtualization overheads and inefficiencies of the modern server architecture to achieve low and bounded tail latencies for lambda functions, and (2) exploiting the right domain-specific architecture (\ie, ASIC-based SmartNICs) to sustain high throughput while reducing CPU load and cost-per-watt in cloud data centers.

%These goals have been echoed repeatedly in both academia and industry.
%For example, projects like DPDK~\cite{dpdk}, Netmap \cite{Rizzo:2012:NNF:2342821.2342830}, and XDP~\cite{xdp} aim at reducing kernel and networking overheads. 
%Works like Dune~\cite{belay2012dune} and Isolate~\cite{cloudflare_isolate} focus on reducing context switching overhead. 
%And, approaches like running CUDA programs on GPUs~\cite{Cook:2012:CPD:2430671} maximize workload throughput by exploiting Single-Instruction-Multiple-Data (SIMD) parallelism.
%However, all such schemes address only a subset of these goals, making them perform poorly for interactive serverless workloads.
%\name{}, on the other hand, is designed to achieve these goals altogehter.

\paragraph{Low latency and high throughput serverless functions.}
The key tenet of serverless compute is that it establishes a clear demarcation between users and infrastructure providers; users only specify programs (or functions) that the providers efficiently execute on their infrastructure. 
Yet, all modern serverless frameworks are based on technologies (\ie, VMs and containers) that were designed to give users explicit control over the underlying infrastructure from the get-go. 
This control---in the form of compute and network (physical and overlay) virtualization---adds a significant overhead to serverless functions.
For interactive serverless workloads, with strict tail latency SLOs, eliminating such computational and networking overheads is becoming crucial~\cite{227649}.

The modern server architecture (with CPUs and GPUs) further adds to these overheads. 
CPUs are Von Neumann machines designed to efficiently execute a long sequence of instructions (or a function). 
However, they perform poorly when executing a large number of small serverless functions where significant time is wasted context switching between functions. 
Similarly, GPUs are Single-Instruction-Multiple-Data (SIMD) machines that serve as look-aside accelerators~\cite{lookaside} in a typical server, controlled by the primary CPU. 
Although, in recent years, these GPUs have shown orders of magnitude improvements in accelerating machine-learning workloads~\cite{Sujeeth:2011:OIP:3104482.3104559, Coates:2013:DLC:3042817.3043086} (which by nature are long-running, batch jobs), they perform poorly for low-latency, interactive tasks (like serverless functions).
Even with technologies (such as GPUDirect RDMA~\cite{GPUDirectRDMA} and RDMA-over-Converged-Ethernet (RoCE)~\cite{rocev2}) that can bypass a CPU and push data directly into the GPU or main memory, the requests for serverless functions still have to traverse a NIC---adding non-negligible delays in the order of sub-microseconds.

\name{} eliminates both these virtualization and architectural overheads by running serverless workloads directly on the vast array of NIC-resident RISC cores.
%, thus achieving bounded tail latency and high throughput while meeting performance SLOs.

\paragraph{A domain-specific processor for serverless functions.}
Till now, cloud providers have almost always relied on newer, faster CPUs to improve applications' performance.
More recently, they have started looking into other fine-grain, domain-specific processors, like look-aside or bump-in-the-wire accelerators (GPUs or FPGAs~\cite{Firestone:2018:AAN:3307441.3307446}). 
This is because, with Moore's Law slowing down~\cite{Hennessy:2019:NGA:3310134.3282307}, CPUs today are no-longer a viable solution to meet ever-rising performance demands of customer workloads in a cost-efficient way---as has been demonstrated by both GPUs (for improving machine-learning training and inference throughput~\cite{Coates:2013:DLC:3042817.3043086}) and FPGAs (for accelerating search indexes~\cite{Putnam:2014:RFA:2665671.2665678} and host networking~\cite{Firestone:2018:AAN:3307441.3307446}).
We believe that ASIC-based SmartNICs present the same opportunity for accelerating serverless workloads with orders of magnitude improvement in performance-per-watt at one-tenth of the hardware cost~\cite{netronome_power, colfax_netronome, Sohan2010Characterizing10G}, compared to server CPUs and GPUs~\cite{Firestone:2018:AAN:3307441.3307446}.
%We give a more detailed cost comparison in \S\ref{subsec:cost_analysis}.

\subsection{Key Challenges}
\label{subsec:challenges}
The embarrassingly-parallel and independent nature of serverless workloads take away much of the complexities that arise when synchronizing state between functions~\cite{Liu:2019:ODA:3341302.3342079}, making them an ideal candidate for SmartNICs with hundreds of cores. 
Still, executing them on these NICs is not a panacea and comes with its own unique challenges:

\paragraph{a. Programming SmartNICs}
Due to their non-cache-coherent design, programming ASIC-based SmartNICs has always been considered hard~\cite{Chen:2005:SAH:1065010.1065038}; non-coherency requires developers to program each NPU core separately, forcing them to manually handle synchronization between individual functions.
With serverless functions, however, this is no longer an issue as functions do not share state and can run independently. 
Still, the lack of an OS layer in these NICs---though useful in reducing unwanted processing---puts the onerous of mapping and placing these functions, across various clusters of cores and memory hierarchy, on the developers; requiring them to have low-level knowledge of the NIC architecture, firmware, and specialized languages it supports.
To take this burden away from developers, we need a high-level abstraction and a framework that can automatically and efficiently compile, optimize, and deploy serverless programs across a collection of these SmartNICs.

\paragraph{b. Offloading serverless workloads}
NPUs are optimized for packet forwarding, and they typically do not support features (\eg, floating-point operations, dynamic-memory allocation, recursion, and reliable transport) that a general-purpose CPU supports~\cite{jungck2011packetc, george2003taming}.
A serverless framework, therefore, must be able to compile workloads that rely on these features by, for example, transforming programs with floating-point operations to fixed-point arithmetic~\cite{Banner:2018:SMT:3327345.3327421}, dynamic-memory allocations to explicit memory calls, recursions to iterations, as well as employing other forms of reliable (or weakly-consistent) delivery protocols (\eg, RoCEv2~\cite{rocev2}, Lightweight Transport Layer~\cite{Caulfield:2016:CAA:3195638.3195647}, or R2P2~\cite{Kogias:2019:RMR:3358807.3358881}).
To achieve high throughput and low latency, emerging workloads are also lowering their dependency on these features. 
For example, deep-learning training and inference is shown to perform well with lower, fixed bit-width integers~\cite{Banner:2018:SMT:3327345.3327421, dettmers20158}.
Moreover, serverless request-response (RPC) pairs are mostly independent and mutually-exclusive, and do not need TCP's strict, reliable, and in-order streaming delivery of messages~\cite{Kogias:2019:RMR:3358807.3358881}. 

\paragraph{c. Ensuring security under multi-tenancy}
Lambdas run alongside other workloads (\eg, microservices) in a data center and share infrastructure resources.
When using SmartNICs: (1) a serverless framework should ensure that lambdas running on the NICs do not interfere with each other or degrade the network performance between the NIC and host CPUs that are running traditional workloads. 
(2) The framework should reserve ample SmartNIC resources (\ie, cores and memory) for basic NIC operations (\eg, TCP/IP offload and checksums) while maximally consuming remaining resources for serverless functions.
(3) Serverless functions should execute in their own isolated sandboxes and the framework should restrict them from accessing each others' working set.
(4) Lastly, the framework should be robust against security attacks (\eg, DDoS) both from outside actors and malicious tenants.

%% file: text/overview.tex
\section{\name{} Overview}
\label{sec:overview}

\name{} adds a new backend to existing serverless frameworks with its own programming abstraction, called \matchlambda{} (\S\ref{subsec:match_lambda}), and the accompanying machine model (\S\ref{subsec:runtime}) that makes it easier to program and deploy lambdas directly on a SmartNIC. 
%We now discuss the programming abstraction and machine model, followed by an implementation of \name{} in \S\ref{sec:implementation}.

\subsection{\matchlambda{} Abstraction}
\label{subsec:match_lambda}

\name{} implements a \matchlambda{} programming abstraction that extends the traditional Match+Action Table (MAT) abstraction~\cite{bosshart13} with more complicated actions (lambdas). 

\paragraph{Programming lambdas}
In \name{}, users provide one or more lambdas written in a restricted C-like language, called Micro-C~\cite{netronome_programming}.\footnote{We use Micro-C as it is the native language of the SmartNICs we have for the evaluation. The Micro-C language can support a large class of serverless functions (\S\ref{subsec:challenges}); however, \name{} is not just limited to this language and can work with more feature-rich languages supported by other SmartNICs.} 
Listing~\ref{lst:lambda_signature} shows the signature of the top-level function, which each lambda must begin with, having two arguments: \verb|headers| and \verb|match_data|. 
The number and structure of all the supported headers (\ie, the \verb|EXTRACTED_HEADERS_T| data structure) and function parameters (\ie, \verb|MATCH_DATA_T|), in \name{}, are defined a-priori. 
The lambdas operate directly on these parameters and headers without having to parse packets, which is done at the parse stage (\Cref{fig:abstract-machine}). 
Furthermore, these functions can have both local objects as well as global objects that persist state across runs.

\vspace{5pt}
\begin{lstlisting}[language={C++}, caption={Signature of the top-level function in Micro-C for the \matchlambda{} abstraction.}, label={lst:lambda_signature}]
int function_name(EXTRACTED_HEADERS_T *headers, MATCH_DATA_T *match_data)
{ 
	// local/global memory and objects.
	return return_value;
}
\end{lstlisting}

Listing~\ref{lst:web_server_lambda} shows a real-world example of a lambda running as a web server. 
The function reads the server address (Line~\ref{lambda_line:6}) from the \verb|headers| variable.
It then copies the requested web content from the memory into the header location pointed by the server address (Line~\ref{lambda_line:8}), before returning.

\begin{lstlisting}[language=C++, caption={An example of a web-server lambda.}, label={lst:web_server_lambda}, escapechar=|]
#define MEM_PER_LAMBDA 20
uint8_t memory[MEM_PER_LAMBDA * 3];
int web_server(EXTRACTED_HEADERS_T *headers, MATCH_DATA_T *match_data)
{
  serverHdr_T *serverHdr =
    hdr_get_serverHdr(headers);|\label{lambda_line:6}|
  memcpy(serverHdr->address, memory,
    MEM_PER_LAMBDA);|\label{lambda_line:8}|
  return RETURN_FORWARD;
}
\end{lstlisting}

\paragraph{Expressing match}
The user further specifies the corresponding P4 code\footnote{We use P4 as it is the most widely used data-plane language~\cite{Bosshart:2014:PPP:2656877.26568900}.} for the match stage (Listing~\ref{lst:generated_match}).
During compilation, the workload manager assigns unique identifiers (IDs) to each of these lambdas, shares this mapping with the gateway, and populates the ID variables (\eg, \verb|WEB_SERVER_ID|, \verb|OTHER_LAMBDA_ID|) in the P4 code.
For each incoming request, the gateway inserts the ID of the destined lambda as a new header. 
The match stage of a \name{} (as defined in the P4 code), checks the ID listed in the new header and calls the matching lambda (implemented as an extern in P4~\cite{p4_spec}) or sends the packet to the host OS, in cases where no matching ID is found.

\vspace{4pt}
\begin{lstlisting}[language=C++, caption={Snippet of a P4 code for the match stage.}, label={lst:generated_match}, escapechar=|]
control ingress {
  if (valid(lambda_hdr)) {
    if (lambda_hdr.wId == WEB_SERVER_ID) {|\label{match_line:3}|
      apply(web_server);
      apply(return_web_server_results);
    } else if (lambda_hdr.wId == OTHER_LAMBDA_ID) {|\label{match_line:5}|
      apply(other_lambda);
      apply(return_other_lambda_results);
    }
  } else { apply(send_pkt_to_host); }
}
\end{lstlisting}

In the end, the workload manager pairs the lambdas (Micro-C code) and match stage (P4 code) into a single \matchlambda{} program, and prepends it with a generic P4 packet-parsing logic. 
It then compiles and transforms this program into a format that the target SmartNIC can execute (\S\ref{sec:implementation}), while ensuring fair allocation of resources and isolation between lambda workloads.

\subsection{Abstract Machine Model}
\label{subsec:runtime}
\begin{figure}[t]
  \centering
  \includegraphics[width=0.9\linewidth]{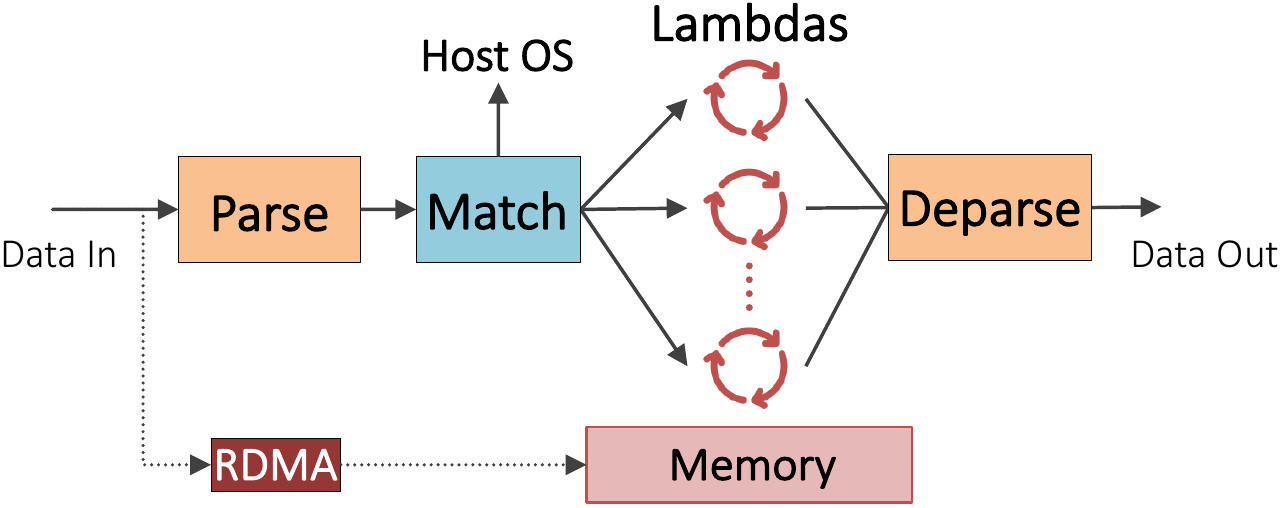}
  \caption{\name{}'s abstract machine model.}
  \label{fig:abstract-machine}
\end{figure}

In \name{}, users write their \matchlambda{} workloads against an abstract machine model (\Cref{fig:abstract-machine}). 
In this model: (1) lambdas are independent programs that do not share state and are isolated from each other; only a matching rule can invoke these functions. 
(2) The match stage serves as a scheduler (analogous to the OS networking stack) that forwards packets to the matching lambdas or the host OS. 
Finally, (3) a parser handles packet operations (like header identification), and lambdas operate directly on the parsed headers. %\ms{should we add a forth point on RDMA.}

These properties of \matchlambda{} machine model make it easier for software developers to express serverless functions by separating out the parsing and matching logic, as well as for hardware designers to efficiently support the model on their target SmartNICs (\S\ref{sec:implementation} demonstrates one such implementation using the Netronome SmartNICs). 
Thus, the abstract machine model enables unique optimizations that lets serverless workloads run as lambdas in parallel without any interference from each other.

\subsubsection{Design Characteristics}
\label{subsubsec:design_specifications}
The abstract machine model has the following three design characteristics:

\paragraph{D1: Run-to-completion execution}
\name{} executes each lambda to completion.
The machine model exposes a dense array of discrete, non-coherent processing threads (\Cref{fig:abstract-machine}), having their own instruction and data store in the memory.
The lambdas execute in the context of a given thread, maximally utilizing the resources of that thread only (\eg, CPU and memory).
Given the short service times and strict memory footprints of serverless functions, modern SmartNICs hold ample resources per thread to execute these lambdas~\cite{netronome_programming}.

Having a large number of parallel threads, further mitigate issues related to head-of-line-blocking where lambdas wait behind other lambadas to finish or context switch, as in the case of server CPUs.
These issues severely affect the performance of lambdas at the tail and require more complicated scheduling policies (like preemption~\cite{kaffes2019shinjuku} or core allocation~\cite{227649}).
With \name{}, however, this is not the case as lambdas---even at the tail---can run to completion without degradation in performance (\S\ref{sec:evaluation}). 

Moreover, the highly-parallel nature and run-to-completion characteristic of the machine model ensure strong \textbf{\em performance isolation} between different lambdas running as separate threads, and \name{} implements weighted-fair-queuing (WFQ)~\cite{wfq} to route requests between these threads.
We leave it as future work to explore more sophisticated resource-allocation mechanisms (\eg, DRF~\cite{Ghodsi:2011:DRF:1972457.1972490}) to further improve the performance of \name{}.

\paragraph{D2: Flat memory access}
The abstract machine model lets users write lambdas assuming a flat (virtual) memory address space.
All objects (local and global) on the thread stack are allocated from within that address space. 
This has the advantage that users do not have to worry about the complex memories, and their structure and hierarchies, present in modern SmartNICs~\cite{netronome_programming}. 
Each of these memories come with their own performance benefits and are necessary to reach high speeds in these NICs; however, having so makes it the responsibility of the programmers to efficiently utilize these memories.
\name{}'s machine model takes this onerous away by exposing a single, uniform memory to the user.

When deploying to a particular SmartNIC, the compiler (or the workload manager) can take into account NIC's specifications and can perform target-specific optimizations to effectively utilize its memory resources (\S\ref{subsec:compiler_optimization}). 
The users can also provide pragmas---specifying which objects are read more frequently---to guide the compiler in allocating objects to memories based on their access needs; it can place small or hot objects to core-local memories, and large or less frequently used ones in external, shared memories.

Furthermore, having a virtual memory space per lambda can let the compiler enforce policies for \textbf{\em data isolation}, since virtual spaces do not interfere and are isolated from each other.
The compiler can insert static and dynamic assertions~\cite{aho1986compilers} to ensure that a lambda does not access the physical memory of other lambdas on the target SmartNIC.

\paragraph{D3: Network transport}
In \name{}, the primary mode of communication between the gateway, lambdas, and external services (\eg, storage) is via Remote Procedure Calls (RPCs).
These RPCs are small, typically single-packet, request-response messages~\cite{spdy, Kogias:2019:RMR:3358807.3358881}.
The parser, in the abstract machine model (\Cref{fig:abstract-machine}), decomposes these messages into headers and forwards them to the match stage, which further directs them to a matching lambda (by looking up the lambda ID associated with each message).
Multi-packet RPCs, depending upon their size, can either be processed directly by the parse and match stage or pushed into the memory over RDMA (\ie, RoCEv2~\cite{rocev2}). 
(In the latter case, an event RPC triggers the lambda to start reading data from the desired memory location.)

Already, modern datacenter applications (like Amazon DynamoDB~\cite{amzn_dynamo} and Deep Learning~\cite{Banner:2018:SMT:3327345.3327421, dettmers20158}) are choosing to go away with the strict, reliable, and in-order guarantees provided by TCP, which are far stronger and computationally intensive than what applications need.
Instead, these applications are being designed to work with weaker guarantees to achieve low tail latency~\cite{Kogias:2019:RMR:3358807.3358881}.
\name{} exploits these facts and assumes a weakly-consistent delivery semantic for RPCs that are processed by the parse and match stage.
A sender (the gateway or external services) tracks the outgoing RPCs to lambdas, and is responsible for resending a message in case of timeouts or packet drops. 
\name{}, on the other hand, performs packet reordering at the SmartNIC for multi-packet RPCs.\footnote{We measured that Netronome SmartNICs can reorder four \SI{100}{B} packets using 120 instructions, which is only 1.3\% of the instructions used by our benchmark lambdas (\S\ref{subsec:optimizer_eval}).}%\ms{add a citation showing how many reorderings are common in a data center.}

%% file: text/implementation.tex
\section{Implementation}
\label{sec:implementation}

\begin{figure}[t]
  \centering
  \includegraphics[width=1\linewidth]{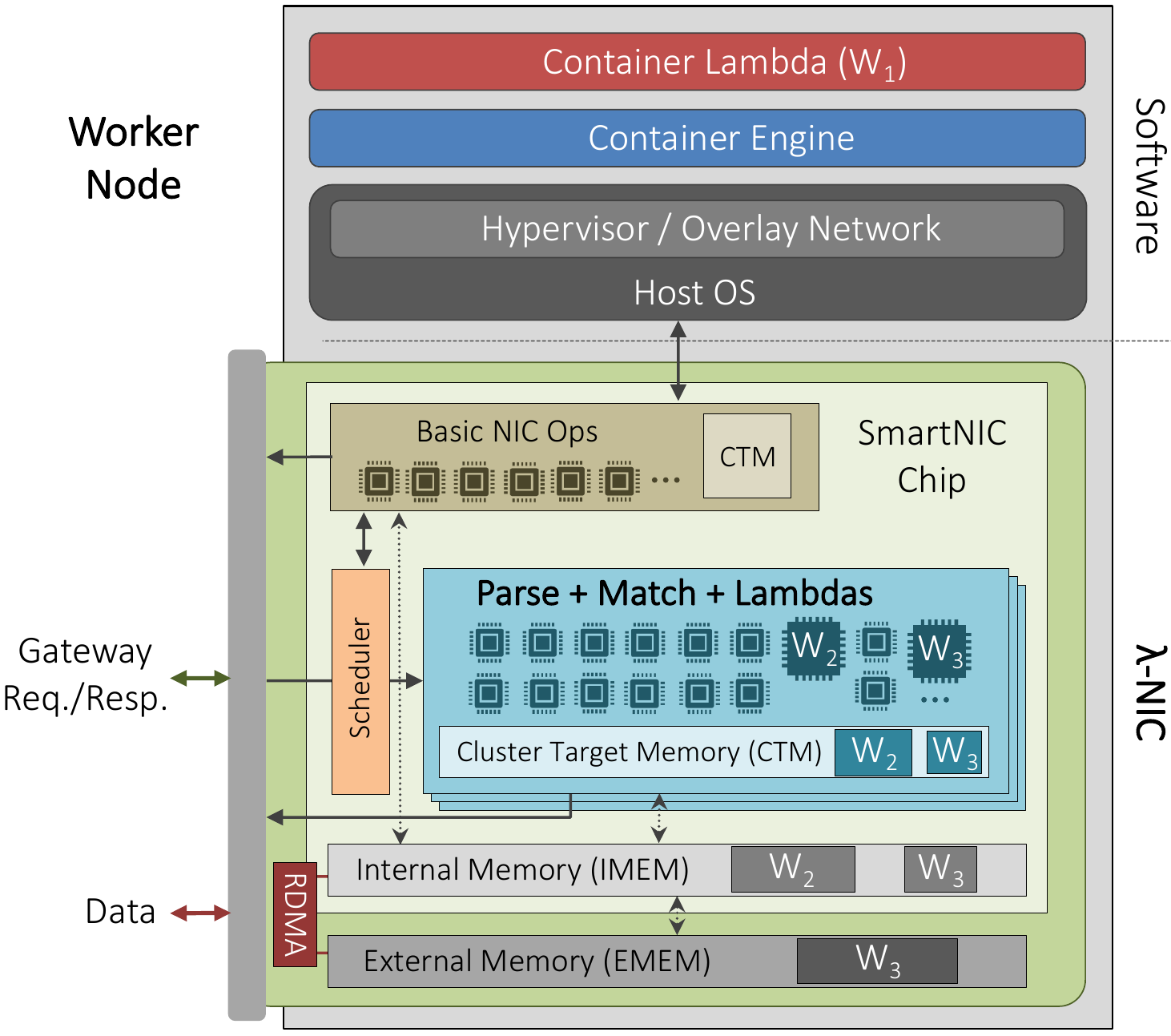}
  \caption{Architecture of a \name{} worker node on a P4-enabled Netronome NIC.}
  \label{fig:runtime_overview}
\end{figure}

In this section, we present an implementation of \name{} on P4-enabled Netronome SmartNICs (\Cref{fig:runtime_overview}).
These NICs contain hundreds of RISC cores grouped together into islands.
Each core has its own instruction and local memory, as well as a Cluster Target Memory (CTM)~\cite{netronome_programming} per island, and is capable of executing multiple threads, concurrently. 
There are also on-chip internal memories (IMEMs) and an external memory (EMEM) shared between all islands and their cores; and a dedicated scheduler unit that directs incoming packets to cores. 
The architecture of Netronome SmartNICs therefore has the necessary elements to efficiently implement \name{}'s \matchlambda{} abstract machine model: cores can execute lambdas to completion (D1), data can reside in different memories (\eg, local, CTM, or more) based on their usage patterns (D2), and the scheduler can direct RPCs to lambdas (D3).%\ms{say something about host communication.}

A more programmable scheduler (\eg, RMT/PIFO-based~\cite{Sivaraman:2016:PPS:2934872.2934899, sp-pifo}) can execute the parse and match stage of the machine model directly, with cores only processing the lambda logic. 
However, the scheduler inside the Netronome NICs we used for our evaluation is not programmable; it is work-conserving and uniformly distributes incoming traffic to all cores. 
We therefore execute all three stages (parse, match, and lambdas) together inside a core, with every core running the same \matchlambda{} program.\footnote{The other approach is to pipeline these stages and run them on separate cores; we intend to look into this as future work.}\footnote{At present, the binary running on SmartNICs must be swapped with a new one each time, resulting in downtimes. However, this constraint is expected to disappear in the next-generation NICs (\S\ref{sec:discussion}).}
For single-packet messages, the scheduler directs an incoming packet to a core at random, which after parsing selects the matching lambda using the ID embedded in the packet. 
The multi-packet messages are committed to memory via RDMA, and lambdas read directly from the desired location.  

\subsection{Target-Specific Optimizations}
\label{subsec:compiler_optimization}

Next, we discuss optimizations, to improve the execution time and binary size of \matchlambda{} programs (\S\ref{subsec:optimizer_eval}), that arise as a result of our design choices and architectural constraints of Netronome NICs.

\paragraph{Lambda coalescing}
As multiple lambdas run on a single core, the workload manager runs program analysis (\ie, dead-code elimination and code motion~\cite{Shahbaz:2015:CIR:2774993.2775000}) to remove duplicate logic (\eg, for modifying similar headers or generating packets) and move it into shared libraries as helper functions. 

\paragraph{Match reduction} 
The workload manager can further reduce the number of tables (\eg, for route-management) defined in the match stage of a \matchlambda{} program. 
Each new lambda consists of both a parse and a match stage; the workload manager can compose these stages with the logic already running on the core, removing the unused headers and duplicate match fields from the final code.
Furthermore, the P4 tables are converted into if-else sequences, which the NIC core can execute more efficiently. 
Transforming tables into if-else sequences also helps reduce the total number of instructions of the final binary, running on a core. 

\paragraph{Memory stratification}
Based on the access patterns, the workload manager can choose the most efficient memory for an object at compile time.
It can also look at the object size or hints from the user (as pragmas) to decide whether to put the object in a local memory, CTM, IMEM or EMEM.

%% file: text/evaluation.tex
\section{Evaluation}
\label{sec:evaluation}
We now compare the performance of \name{} with bare-metal and container backends both in isolation, when executing a single lambda (\S\ref{subsubsec:isolation}), and in a shared setting, when running multiple lambdas together (\S\ref{subsubsec:contention}). We also evaluate the impact of \name{} on resource utilization, startup times, as well as the effectiveness of the \name{} compiler to optimize lambda program size when running on the NIC cores (\S\ref{subsec:other_metrics}).

\subsection{Test Methodology}
\label{subsec:methodology}

\subsubsection{Baseline framework}
\label{subsubsec:framework}
To evaluate \name{} against existing serverless backends, we select OpenFaaS~\cite{openfaas} as our baseline serverless compute framework, which closely resembles the architecture depicted in \Cref{fig:serverless_overview}.
We choose OpenFaaS for its simplicity, ease of deployment, extensive feature set, and greater adoption by the community.\footnote{OpenFaaS is the most favorable open-source serverless compute framework with $\sim$12,000 stars on GitHub~\cite{serverless_comparison}.}
It is written in Golang~\cite{golang} and includes: (1) a Web UI, (2) an autoscaler to scale lambdas as demands change, (3) a Prometheus~\cite{prometheus} based monitoring engine to analyze system state and (4) a gateway with a NAT~\cite{rfc2663} to proxy users' requests to the appropriate lambdas.
Each of these components and lambdas run as Docker~\cite{docker} containers, managed via Kubernetes~\cite{kubernetes} or Docker Swarm~\cite{docker_swarm}. 

\begin{figure}[t]
  \centering
  \includegraphics[width=0.85\linewidth]{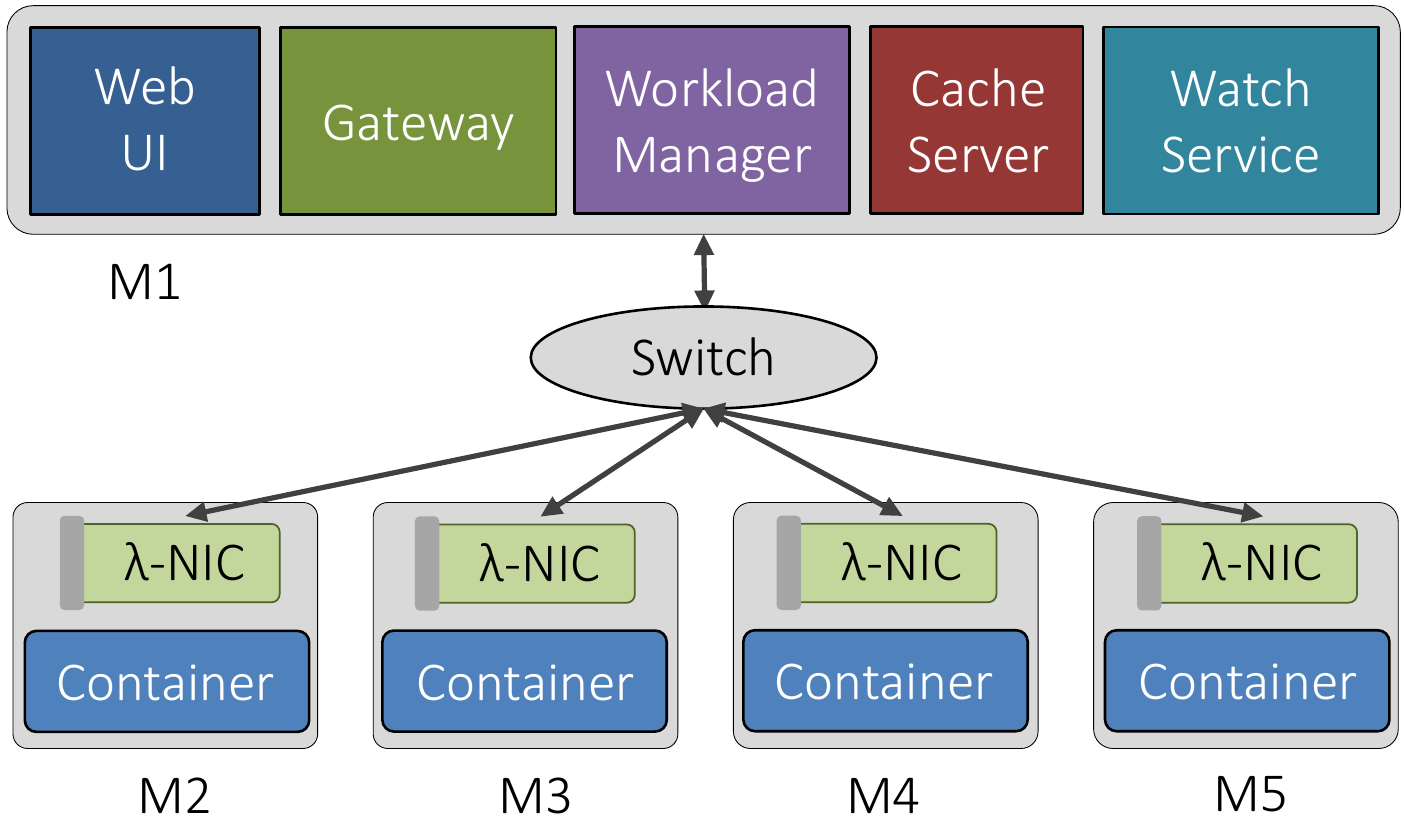}
  \caption{An overview of the testbed having 4 worker nodes and a master node all connected to a \SI{10}{G} switch.}
  \label{fig:experiment_overview}
  \vspace{5pt}
\end{figure}

\begin{figure*}[t]
    \centering
    \includegraphics[width=\linewidth]{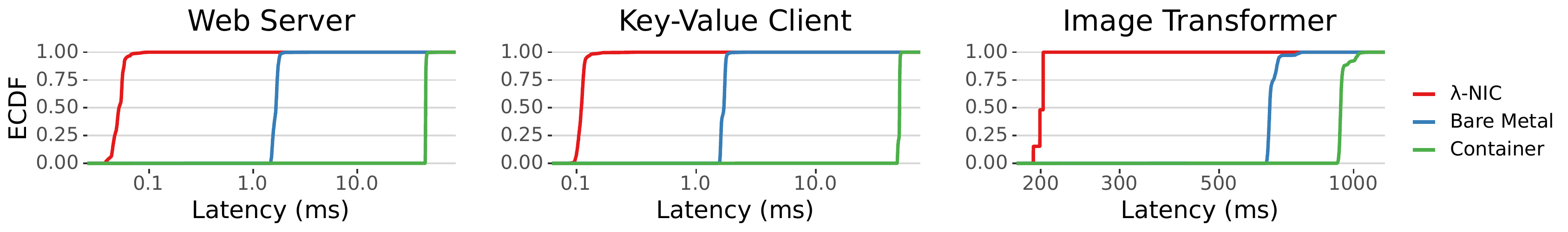}
    \caption{ECDF of latencies, in log scale, when executing a single workload instance in isolation.}
    \label{fig:cdf_latency}
\end{figure*}

\begin{figure*}[t]
  \centering
  \includegraphics[width=\linewidth]{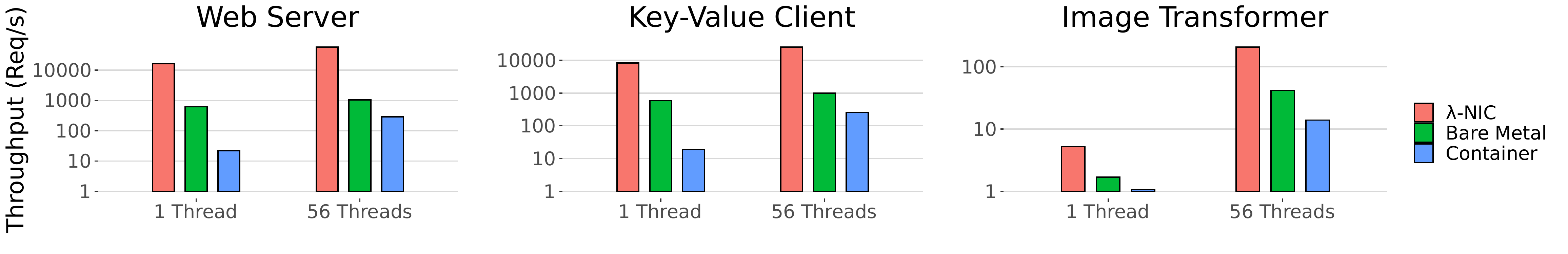}
  \vspace{-20pt}
  \caption{Average throughput, in log scale, when executing a single workload instance in isolation. (Standard deviation is negligible across all runs.)}
  \label{fig:avg_throughput}
  \vspace{-10pt}
\end{figure*}

\paragraph{Adding a bare-metal backend}
For evaluating emerging runtimes like Isolate~\cite{cloudflare_isolate}, we add support for a bare-metal backend to OpenFaaS.
It is implemented as a Python service\footnote{We found the performance of the Python service similar to a comparable C implementation except for the one-time startup cost of the backend. We, therefore, used the Python service for its ease of integration with OpenFaaS.} that runs on a bare-metal server as a standalone process, launching lambdas as new threads to serve users' requests.
The service relies on a Raft-based~\cite{Ongaro:2014:SUC:2643634.2643666} distributed key-value store, called etcd~\cite{etcd}, to sync lambda-related states (\eg, number of active lambdas, their placement and load balancing policies) with the gateway to correctly proxy requests.  
Our goal, using the bare-metal backends, is to analyze how performance of lambdas improves in the absence of the container processing stack.

\paragraph{Introducing \name{} extensions}
We built \name{} as an extension to our baseline framework, inheriting all of OpenFaaS's core features with additional support for running lambdas on P4-enabled Netronome SmartNICs.
With these extensions, the baseline framework can simultaneously deploy lambdas to containers, bare-metal, and SmartNIC backends.
We also augment etcd to share state and manage \name{} deployments across multiple worker nodes.%\ms{add details as to why we have not use a match-action table in our NICs.}

\subsubsection{Testbed setup} Our evaluation testbed consists of a cluster of five servers (\Cref{fig:experiment_overview}) housing two Intel Xeon Gold 5117 processors with 14 physical cores, running at \SI{2.0}{GHz} with \SI{32}{GiB} DDR4 \SI{2666}{MT/s} Dual-Ranked RAM and \SI{120}{GiB} SATA SSD.
One of the servers (M1) act as a master node running: Kubernetes services, gateway, workload manager, memcached server~\cite{memcached}, the web interface, and the monitoring engine.
M1 comes with a Broadcom 57412 2x\SI{10}{Gb} and 2x\SI{1}{Gb} Quad-Port NIC, which is used for management traffic.
Other servers (M2--5) are worker nodes equipped with a Netronome Agilio CX 2x\SI{10}{Gb} SmartNIC~\cite{netronome_cx} having 56 RISC cores (8 threads and \SI{16}{K} instructions per core) running at \SI{633}{MHz} with \SI{2}{GiB} of on-board RAM.
All servers connect to an Arista DCS-7124S switch over a \SI{10}{Gbps} link. 
The backends communicate over an overlay network using Kubernetes' calico~\cite{calico} networking plugin for high performance switching and policy management across nodes inside a Kubernetes cluster.

\subsection{Benchmark Workloads}
\label{subsec:workloads}
We evaluate \name{} on three different types of interactive lambdas (\ie, a web server, a key-value client, and an image transformer), each reflecting a popular use case~\cite{azure_serverless_cookbook, lambda_use_cases, goog_cf_usecases}.

\paragraph{a. Web server}
A common usage pattern for lambdas is to serve web contents~\cite{lambda_use_cases}, such as text or HTML pages, similar to traditional web servers (like nginx~\cite{nginx}).
These workloads are typically self-contained and do not need information from external sources (\eg, data stores) to service a request.
For our experiments, we wrote a lambda that returns text responses based on the incoming requests.

\paragraph{b. Key-value client}
Next, we consider workloads with external dependencies, needing information from remote services.
These workloads query users' data from external storage, \eg, databases or key-value stores (such as memcached~\cite{memcached}), do customization on the retrieved data, and finally send the processed data back to the user.
Moreover, these workloads generate extensive intra-data center requests and typically have strict tail-latency requirement to meet user service-level objectives (SLOs).
To evaluate, we implement lambdas acting as key-value clients that generate write (SET) and read (GET) requests to a memcached server.

\paragraph{c. Image transformer.}
Finally, we evaluate workloads that involve real-time, interactive processing of large datasets (\ie, image processing or stream processing)~\cite{Viola:2004:RRF:966432.966458, Chen:2016:RDP:2882903.2904441}, where the datasets span multiple packets and must be stored in memory (\ie, DRAM).
These workloads perform customization to the requested datasets, and either return a response to the user immediately or store results back to the memory for further processing~\cite{lambda_image_editing}.
For our evaluation, we consider lambdas that transform RGBA images to grayscale.

\subsection{System Performance}
\label{subsec:performance}
We now discuss how \name{} performs, both in terms of latency and throughput, compared to the bare-metal and container backends. We evaluate two cases: (1) when there is only a single lambda running on a backend in isolation, and (2) when there are multiple lambdas running, all contending for the shared resources (\ie, processing cores and memory).

\subsubsection{Performance in Isolation}
\label{subsubsec:isolation}
We first look at the latency and throughput of a lambda in isolation.

\paragraph{Latency}
We measure the latency of each backend, which is the time it takes for a gateway to send a request to a node running a single thread of the pre-loaded (or warm) lambda and receive a response back (\Cref{fig:cdf_latency}).
For web server and key-value client lambdas, \name{} outperforms containers by 880x and bare-metal by 30x in average latency---completing requests in under \SI{100}{ns}---while still achieving 5x to 3x improvements for the data-intensive image-transformer lambda. 
Improvement are more visible at the tail (\ie, 99th-percentile) where \name{} achieves 5x to 24x better tail latency than bare-metal for the three benchmark lambdas.
For the key-value client lambdas, \name{} even improves upon reported latencies in a highly-optimized cloud-scale data center~\cite{fb_memcached} by three orders of magnitude.
Furthermore, container and bare-metal exhibit longer tail latency, specifically, for short-lived web server and key-value lambdas.
This is likely the artifact of miscellaneous software overheads (\eg, context switching, cache management, and network stack).

\paragraph{Throughput}
We see similar improvements in throughput, measured as request serviced per second, for the three lambdas when running on \name{}.
We carryout two separate experiments: (1) closed-loop testing with sender generating each request one after the other, and (2) parallel testing with 56 requests---the maximum number of threads that can run simultaneously on our testbed server CPU---to stress the backends under concurrent load.
\name{} outperforms both container and bare-metal backends (\Cref{fig:avg_throughput}), servicing requests 27x to 736x faster than the two backends for the web server and key-value client lambdas, and 5x to 15x faster for the image-transformer lambda.

\subsubsection{Performance Under Resource Contention}
\label{subsubsec:contention}
In a real setting, a serverless backend will typically run multiple lambdas at the same time.
These workloads will contend for their fair share of resources (\ie, CPU and memory), leading to added delays due to, for example, context switching of lambdas and movement of data to and from CPU and memory.

\begin{figure}[t]
  \centering
  \includegraphics[width=\linewidth]{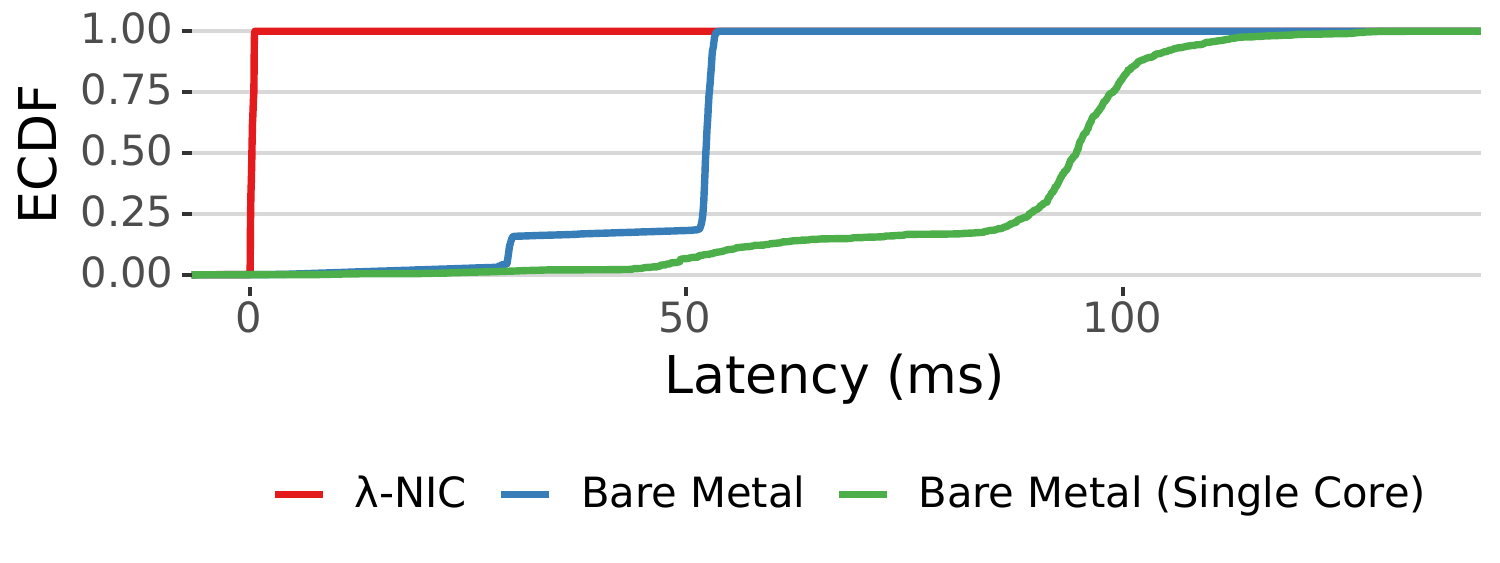}
  \vspace{-20pt}
  \caption{CDF of latencies when running three distinct web server lambdas concurrently.}
  \label{fig:cdf_context_switch}
\end{figure}

\begin{table}
	\small
    \begin{tabular}{ c r r r }
    \toprule
                                        &
    \multirow{2}{*}{\bfseries \name{}}  &
    \multicolumn{2}{c}{\bfseries{Bare Metal}} \\
    \cline{3-4}
                                &
                                &
    {\bfseries 56 Threads}        &
    {\bfseries 1 Thread}          \\
    \midrule
    {\bfseries Throughput (Req/s)} &
    58,000                         &
    950                            &
    520                           \\
    \bottomrule
    \end{tabular}
    \caption{Average throughput when running three distinct web server lambdas concurrently. (Standard deviation is negligible across runs.)}
    \label{tab:wc_context_switch}
\end{table} 

\paragraph{Effects of context switching}
In the previous experiment, we measured the performance of each backend when running a single lambda in isolation.
Now, we evaluate a set up with three distinct web-server lambdas running on a single backend at once. 
We generate requests for each of these workloads in a round-robin fashion, causing the processor to context switch between lambdas when servicing each incoming request.
\Cref{fig:cdf_context_switch} shows the context-switching overhead on latency for \name{} as well as the bare-metal backend (with 1 and 56 threads) when executing the warm (pre-loaded) web-server lambdas. 
With multiple lambdas running concurrently, the bare-metal backend suffers even higher latency (178x to 330x) compared to \name{}.
Moreover, \name{} completes requests 55x to 100x faster than the bare-metal backend, with both single and 56 threads (\Cref{tab:wc_context_switch})---a difference of at least 9\%  compared to running a single lambda in isolation (\S\ref{subsubsec:isolation}), whereas \name{} shows no significant change.
The performance of containers was a lot worse than even the bare-metal backend (not shown).
These experiments demonstrate that, unlike container and bare-metal backends (with server CPUs), \name{} is not susceptible to context switching and performs better under resource contention by virtue of a vast array of on-chip NPU cores and the lack of operating system and container software.

\subsection{Other Metrics}
\label{subsec:other_metrics}

\paragraph{Resource utilization}
We also compare the memory and CPU usage at the host and the SmartNIC when running a single data-intensive image-transformer instance in isolation.
\Cref{tab:resource_usage} shows the additional resources utilized by each backend when servicing 56 concurrent requests.
Containers have the largest memory footprint and consume an order of magnitude more host memory and CPU cycles than the bare-metal backend.
On the other hand, as expected, \name{}'s impact on the host memory and CPU is negligible, and it consumes roughly the same amount of NIC memory for the image-transformer workload as the bare-metal backend on the host.
%\ms{Sean, we should say something about how memory is not a linear function of number of workloads you have on the NIC.}, which is the minimum workload logic needed to process requests.

\begin{table}
    \small
    \begin{tabular}{ l r r r}
    \toprule
                            &
    {\bfseries \name{}}     &
    {\bfseries Bare Metal}  &
    {\bfseries Container}   \\
    \midrule
    {\bfseries Host CPU {\footnotesize (Avg. \%)}} &
    +0.1                           &
    +9.2                           &
    +13.7                \\
%    \hline
    {\bfseries Host Memory {\footnotesize (MiB)}}  &
    0	                       &
    +62.5                     &
    +219.5                \\
%    \hline
    {\bfseries NIC Memory {\footnotesize (MiB)}}  &
    +63.2                &
    0                    & 
    0                       \\
    \bottomrule
    \end{tabular}
    \caption{Additional resources utilized by each serverless backend for the image-transformer workload.}
    \label{tab:resource_usage}
%    \vspace{-10pt}
\end{table}

\begin{table}
	\small
    \begin{tabular}{ l r r r}
    \toprule
                                &
    {\bfseries \name{}}         &
    {\bfseries Bare Metal}      &
    {\bfseries Container}       \\
    \midrule
    {\bfseries Workload Size ({\footnotesize MiB})} &
    11.0                         &
    17.0                         &
    153.0                        \\
%    \hline
    {\bfseries Startup Time (s)} &
    19.8                         &
    5.0                          &
    31.7                         \\
    \bottomrule
    \end{tabular}
    \caption{Factors affecting startup times.}
    \label{tab:boot_up}
    \vspace{-5pt}
\end{table}

\paragraph{Startup times}

Next, we measure the startup time; the total time a backend takes to download the image-transformer binary and its dependencies and start serving requests. 
To compare startup times we use: (1) Lambda binary size (\ie, SmartNIC's compiled firmware, Python library packaged using setuptools and Wheel~\cite{python_wheel}, and the docker container). (2) Boot-up time of the lambda, from launching the system to responding to a user request.
\name{}'s image-transformer binary is 13x smaller in size than a container image, and is comparable to a bare-metal binary (\Cref{tab:boot_up}). 
In addition, it takes the image-transformer lambda 38\% less time on \name{} to service the first request compared to a container.
On the bare-metal backend, the image-transformer binary starts up in under 5 seconds (4x faster than \name{}); however, in our evaluation, we do not consider any framework overheads (like Isolate~\cite{cloudflare_isolate}), which will likely lead to higher startup times.
In summary, while slow start is a known issue in serverless compute, \name{} keeps the additional delay over bare-metal backends 2x less than the container overhead.

\paragraph{Optimizer effectiveness}
\label{subsec:optimizer_eval}

We now report the results of our compiler optimizations.
The number of instructions in the na\"ive implementation---consisting of two key-value clients, a web server, and an image transformer lambda---is gradually reduced by applying the following target-specific optimizations (\S\ref{subsec:compiler_optimization}). 
First, we perform {\em lambda coalescing} for the two distinct key-value clients.
We coalesce these lambdas, as they contain equivalent logic to generate a new packet to query memcached, which we can combine and reuse.
We further coalesce the web server and image-transformer lambdas, having a pattern of response that does not query external services.
Hence, we combine their reply logic.
Next, we apply {\em match reduction}. 
The na\"ive implementation adds a separate table for managing routes for each lambda.
We combine these tables into one, and use individual parameter values (defined as P4 metadata) for route management.
Finally, we do {\em memory stratification} to place variables into appropriate memories based on their sizes.
For example, the image variable within the image-transformer lambda is mapped to IMEM, whereas the web server results are mapped to CTM inside the island.
These optimizations bring the total number of instructions of the final binary down to 8,050 (a reduction of 9.56\% from the na\"ive implementation); hence, improving latency by \SI{6.3}{\mu s} (on average)~\cite{Crowley:2000:CPA:335231.335237} or letting additional lambdas to fit within the program-size constraints of the Netronome SmartNIC.

\begin{figure}[t]
  \centering
  \includegraphics[width=0.95\linewidth]{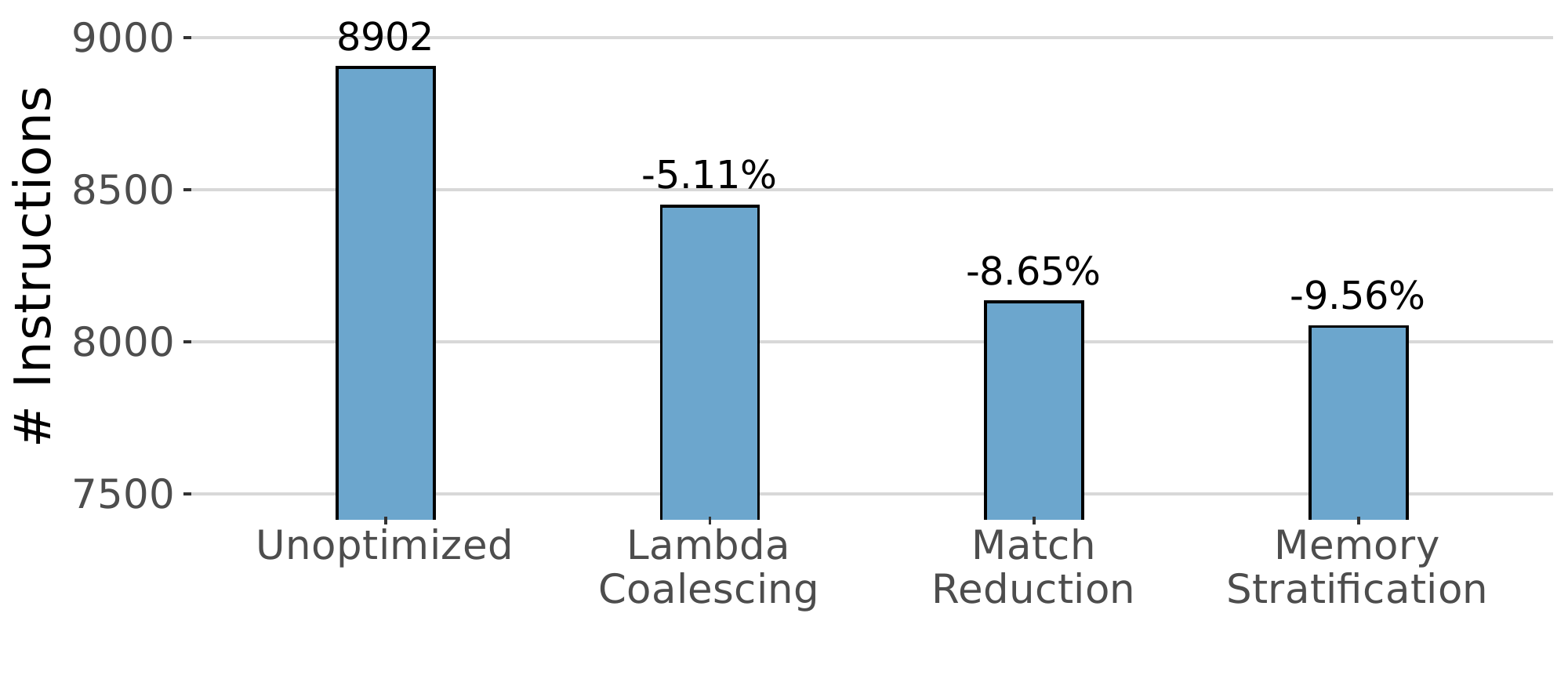}
  \vspace{-15pt}
  \caption{Effectiveness of \name{}'s target-specific optimizations in reducing the code size.}
\end{figure}

\iffalse
\subsection{Infrastructure Cost Analysis}
\label{subsec:cost_analysis}
We now discuss the difference in deployment cost of \name{} versus
existing frameworks.
At the time of writing, a cheapest physical core with two hyper threads sells for about \$0.051/hr\footnote{This is the hourly cost of an AWS EC2 a1.large instance.}, which is roughly \$450/yr and \$2250 over the lifetime of a server---given that a server typically lasts around 3 to 5 years in a data center~\cite{Firestone:2018:AAN:3307441.3307446}.
Similarly, buying 6 more CPU cores alone costs around \$1,944 or around \$324 per core (comparing 20 core Intel Xeon 6138 versus 14 Core Intel Xeon 5117) over the lifetime of a
server.
In comparison, 10~Gb Netronome Agilio with 56 cores and 2~GiB of RAM, which is used for our testbed, sells for \$444~\cite{colfax_netronome}. 
Given that a traditional 10G NIC sells for \$100, replacing it with a SmartNIC adds \$344 over 5 years.
Thus, if 56 NPUs on a SmartNIC are used in a similar fashion as a single CPU core, the saving is around \$1906 per server over its lifetime.
Furthermore, since SmartNIC houses 2+~GiB of RAM, a SmartNIC not only gets us more cores but also more memory.
As for the power consumption, Netronome Agilio consume less than 25~W~\cite{netronome_power}, which is on par with traditional NICs~\cite{Sohan2010Characterizing10G}.
Adding 6 cores incurs about 20~W extra power, or 3~W added power per core.
This shows that \name{} gives substantial cost savings compared to the CPU counterparts.
\fi

%% file: text/discussion.tex
\section{Discussion}
\label{sec:discussion}

\paragraph{Choice of hardware}
\name{} is not just limited to NPU-based SmartNICs.
In fact, the \name{}'s abstract machine model can run on other SmartNICs (with varying benefits) having more general-purpose processors: either FPGA- or SoC-based SmartNICs, or some form of ASIC with ARM cores~\cite{netronome_fx}.
These alternatives can further extend the processing capabilities of \name{}, providing support for more features (such as floating point operations, deeper instruction store, and dynamic memory allocation) to run more complicated workloads.

\paragraph{Hot swapping workloads}
For each new lambda, \name{} needs to recompile and swap the firmware with the one currently running on the SmartNIC.
Present versions of Netronome SmartNICs do not support hot swapping or hitless updates~\cite{netronome_programming}, resulting in downtimes each time a new firmware is loaded.
Hitless updates are not technically challenging as devices like FPGAs~\cite{xilinx_fpga_reconfig} and programmable switch fabrics~\cite{bosshart13} already support it (using partial reconfiguration and versioning techniques).
We believe this limitation will go away in the future versions of SmartNICs as well, allowing \name{} to load new lambadas with causing downtimes.

\paragraph{Accelerating other forms of workloads}
In this work, we primarily focused on offloading interactive serverless workloads to SmartNICs.
However, \name{} can accelerate other parts of the serverless framework as well.
For example, the gateway is a proxy that routes users' requests to \name{}'s worker nodes (\Cref{fig:serverless_overview}); its performance is therefore crucial to the end-to-end behavior of the system.
\name{} can provide strict bounds on tail latency and throughput, by running the gateway directly on a SmartNIC.
Certain types of data stores (like key-value stores~\cite{memcached}) can also benefit from \name{}.
Their restricted compute pattern~\cite{Jin:2017:NBK:3132747.3132764} lends itself nicely to run on \name{}'s \matchlambda{} machine model.
Likewise, existing large and long-running workloads (like code compilation and video processing) have shown to perform better when broken down into small serverless functions~\cite{fouladi2017thunk}; thus, making \name{} a suitable target for these workloads.
We believe that these trends will inspire more workloads to migrate to \name{} in the future, with serverless frameworks automatically determining which backend to execute the workload on. 

\paragraph{Security and reliable transport}
The serverless framework (\ie, gateway, workload manager, and worker nodes) typically run within a trusted domain of a provider or a tenant, such that any malicious
attempt to trigger the lambdas will be blocked by the gateway. 
In addition, each lambda operates within its own memory region on the NIC, restricting them from accessing each others data.
\name{} enforces this policy using compile-time assertions; in the future, we plan to explore enforcing assertions at runtime for dynamically-allocated memory (once it becomes available in the upcoming SmartNICs).
For transport, \name{} relies on RDMA for multi-packet messages. 
However, with recent focus on terminating the entire transport layer on the NIC~\cite{Montazeri:2018:HRL:3230543.3230564, tonic}, \name{}'s serverless functions can instead operate on complete messages rather than individual packets.

%% file: text/related-work.tex
\section{Related Work}
\label{sec:related}

\paragraph{NIC offloading technologies}
Offloads for network features, such as L3/L4 checksum computation, large send and receive offload (LSO, LRO)~\cite{lso}, and Receive Side Scaling (RSS)~\cite{rss} have been around for decades.
More recent work, however, is looking into offloading more complex, application-related functions to modern programmable, SmartNICs~\cite{Firestone:2018:AAN:3307441.3307446}.
For example, Microsoft is using FPGA-based SamrtNICs to offload hypervisor switching tasks~\cite{Firestone:2018:AAN:3307441.3307446}, which were previously handled by CPU-based software switches (like Open vSwitch (OVS)~\cite{ovs}).
HyperLoop~\cite{Kim:2018:HGN:3230543.3230572} provides methods for accelerating replicated storage operations using RDMA on NICs. 
\name{} can assist these works by providing a framework to easily deploy general compute on a cluster of nodes hosting SmartNICs.
Both Floem~\cite{floem} and iPipe~\cite{Liu:2019:ODA:3341302.3342079} provide a framework to enable easier development of NIC-assisted applications.
However, these frameworks can offload only a portion of these applications to a NIC as a bump-in-the-wire, and need CPUs to do the remaining processing.
In contrast, \name{} runs complete workloads on the NIC, mitigating the effects of any CPU-related overheads.

\paragraph{In-network computing}
Orthogonal to NIC offloading technologies, there is a recent focus on moving various application tasks inside the network. 
P4~\cite{Bosshart:2014:PPP:2656877.26568900} and RMT~\cite{bosshart13} have provided the initial building blocks: a data-plane language and an architecture for programmable network devices, which enabled developers to run various applications in-network.
For example, SilkRoad~\cite{Miao:2017:SMS:3098822.3098824} and HULA~\cite{Katta:2016:HSL:2890955.2890968} present methods for offloading load balancers, NetCache~\cite{Jin:2017:NBK:3132747.3132764} implements a key-value store, and NetPaxos~\cite{Dang:2015:NCN:2774993.2774999} runs the Paxos~\cite{Lamport:1998:PP:279227.279229} consensus algorithm inside switches.
Tokusashi et al.~\cite{Tokusashi:2019:CIC:3302424.3303979} further demonstrate that in-network computing not only improves performance but also is more power efficient.
\name{}, alongside these networking devices, can provide a more programmable environment for accelerated application processing. 
In fact, SmartNICs have more memory and less-restricted programming model, which can help alleviate the limitations present in these switches.

\paragraph{Improving serverless compute}
Serverless compute is a relatively new idea and many of its details are not yet disclosed by the cloud providers.
Thus, most of the recent work focuses on reverse engineering existing frameworks to study their internals and to educate the public.
For example, OpenLambda~\cite{openlambda} provides an open-source serverless compute framework that closely resembles the ones deployed by the cloud providers.
Glikson et al.\cite{Glikson:2017:DEC:3078468.3078497} proposed another framework with support for edge deployments. 
\name{} complements these efforts by presenting a high-performance, open-source serverless compute framework for testing and developing lambdas on SmartNICs.

%% file: text/conclusion.tex
\section{Conclusion}
\label{sec:conclusion}
Server CPUs are not the right architecture for serverless compute.
The particular characteristics of the severless workloads (\ie, fine-grain functions with short service times, and memory), as well as the slowdown of Moore's Law and Denard Scaling, demand a radically different architecture for accelerating serverless functions (or lambdas): an architecture that can execute multiple of these lambdas in parallel with minimal contention and context-switching overhead.
In this paper, we present \name{}, a serverless framework along with an abstraction, \matchlambda{}, and a machine model for executing thousands of lambdas on an ASIC-based SmartNIC.
These SmartNICs host a vast array of RISC cores with their own instruction store and local memory, and can execute embarrassingly parallel workloads with very low latency.
\name{}'s \matchlambda{} abstraction hides the architectural complexities of these SmartNICs and efficiently compiles, optimizes, and deploys serverless functions on these NICs---improving average latency by 880x and throughput by 736x.
With new and emerging developments in both SmartNICs and serverless frameworks, we believe offloading lambdas to SmartNICs will become a common practice, inspiring even more complex real-world workloads to run on \name{}.